\documentclass[a4paper,11pt]{article}
\usepackage{pos}

\usepackage{lineno}

\title{Luminosity Calibration at the CMS Experiment}

\author*[a,b]{Attila József Rádl}

\onbehalf{for the CMS collaboration}

\affiliation[a]{HUN-REN Wigner Research Centre for Physics,\\
  Konkoly-Thege Miklós út 29-33, Budapest, H-1121, Hungary}

\affiliation[b]{Eötvös Loránd University,\\
Pázmány Péter sétány 1A, Budapest, H-1117, Hungary}

\emailAdd{attila.radl@cern.ch}
\emailAdd{radl.attila@wigner.hun-ren.hu}

\abstract{Precision luminosity measurements are essential to determine the fundamental properties of the physics processes at the LHC. The estimation of the integrated luminosity at the CMS experiment requires absolute scale calibration of the luminometers, which is derived under special LHC machine setup. Series of beam separation (van der Meer) scans are performed during these special runs. The transverse profile of the overlap between the proton beams are estimated by the continuous monitoring of the interaction rates together with the beam properties. The dominant sources of systematic calibration uncertainty are related to the precise estimation of the beam separation and the non-factorizability of the proton density distributions in the transverse direction. The correction factors and their uncertainties are extracted for each effect and propagated to determine the final absolute scale and the corresponding uncertainty. The obtained van der Meer scan calibration is applied to the full physics data-taking period in order to estimate the integrated luminosity. The latest results of the luminosity calibration studies are reported from the CMS experiment.}

\FullConference{The Eleventh Annual Conference on Large Hadron Collider Physics (LHCP2023)\\
 22-26 May 2023\\
 Belgrade, Serbia\\}

\begin{document}
\maketitle

\section{Luminosity Calibration Strategy at the CMS}
\noindent The Compact Muon Solenoid (CMS) \cite{CMS:2008xjf} experiment is designed to explore the fundamental parameters and the region of validity of the standard model. Luminosity estimation is a key ingredient for precision cross section studies. Luminometry is based on the continuous monitoring of the event rate ($dN/dt$) by various detectors (luminometers). The instantaneous luminosity ($\mathcal{L}$) is directly proportional to this rate via the cross-section ($\sigma$) of the associated process:
\begin{equation}
    \frac{dN}{dt} = \mathcal{L} \cdot \sigma.
    \label{lumi}
\end{equation}

The absolute luminosity scale is determined in dedicated transverse beam separation, so called van der Meer (vdM) scans~\cite{vanderMeer:1968zz}. These scans are carried out in unique low-pileup conditions and with low-intensity colliding bunches, which are widely separated in time to prevent cross-talk between the adjacent crossings. Precise calibration requires separating the two beams up to $\pm 6~\sigma_{\rm{beam}}$ typically in 25 steps. Together with the beam separation, the event rates are continuously monitored to determine the calibration constant that is specific to the detector. This constant is known as the visible cross-section ($\sigma_{\rm{vis}}$):
\begin{equation}
    \sigma_{\rm{vis}} = \frac{2 \pi \Sigma_x \Sigma_y R_0}{N_1 N_2 f},
    \label{overlap_integral}
\end{equation}
where $\Sigma_{x/y}$ represent the beam overlap widths in x/y transverse directions assuming that the proton density function can be factorised into two independent transverse terms depending only on x and y, $N_{1/2}$ correspond to the number of particles in the two bunches, while $f$ is the revolution frequency of the LHC, and $R_0$ is the measured head-on rate.

The vdM calibration results are extrapolated to regular data-taking periods assuming that the same type of particles are colliding with identical centre of mass energy. In this scenario the $\sigma_{\rm{vis}}$ is not expected to change. However, the significant rise of the pile-up and presence of bunch trains introduce potential biases in the luminometer measurements.  For example, out-of-time signals originating from previous bunch crossings can be produced by spill-over of the electronic signal or late particles or the deactivation of the detector material. After adjusting the rates for these effects, relative corrections for time dependent efficiency as well as linearity are derived from short vdM-like emittance scans during physics runs.

\section{Luminosity Detectors}
\noindent The Beam Radiation Instrumentation \& Luminosity (BRIL) project is responsible for luminosity instrumentation and real-time measurement. The luminometers are expected to have a high-level of long-term stability, and linear signal-luminosity characteristics across diverse conditions. 

Several detectors are capable of measuring the luminosity bunch-by-bunch. Two dedicated luminometers are installed to monitor the event rate efficiently. The Pixel Luminosity Telescope (PLT) consists of three pixel planes in a telescope arrangement, while the Fast Beam Condition Monitor (BCM1F) consisted of silicon and diamond sensors mounted on the same C-shape holder in Run 2 (and upgraded to actively cooled Si-sensors for Run~3)~\cite{Hayrapetyan:2870088}. Beside providing real-time luminosity, BCM1F is able to measure the machine-induced background as well. Two rings of the hadron forward calorimeters with dedicated back-end read out are employed as luminometers with two algorithms implemented: the HFET algorithm computes the sum of the transverse energy deposits, and the HFOC algorithm counts calorimeter towers with energy deposits exceeding a particular threshold. The occupancy of the pixel detector changes linearly with luminosity. The pixel cluster counting algorithm (PCC) estimates the number of clusters (corresponding to energy deposits by charged particles passing through the detector layers) using online reconstruction in the high-level trigger. The muon barrel drift tubes (DT) count muon track stubs integrated over the full orbit during Run~2. Furthermore, ionization chambers from the CERN radiation monitoring system for the environment and safety (RAMSES) measure the ambient dose equivalent rate due to ionizing radiation integrating over periods of about 1 second. 

\section{Luminosity Precision for Run~2}
\noindent CMS recorded during the four years of Run~2 a data set corresponding to almost 140~fb$^{-1}$ of integrated luminosity. The overall uncertainty of the luminosity estimation includes both the calibration-related sources that affect the determination of $\sigma_{\rm{vis}}$ and the integration-related sources arising from the extrapolation to nominal physics conditions as well as time stability. The luminosity values together with the corresponding uncertainties are measured for each data taking period (collision type, energy and year) separately. The most precise offline luminosity is obtained for 2016 pp collisions at 13 TeV with only 1.2\% uncertainty~\cite{CMS:LUM-17-003}. The accuracy is limited due to some dominant systematical effects, details are provided in Table~\ref{uncertainties}. 

\paragraph{Calibration Uncertainties}
LHC provides nominal beam displacements defined by the configurations of the steering magnets. However, calibration of the absolute beam separation requires comparison between nominal positions and CMS tracker coordinate system via the position of offline reconstructed vertices. Length-scale corrected beam displacements are then used to calculate the beam overlap integrals.

The orbit of each beam is derived from the measurements of the beam position monitors (BPMs). The Diode Orbit and Oscillation (DOROS) devices are located near the CMS, while the arc BPMs are installed in the LHC arc adjacent to the interaction point. The positions used in the overlap integral are acquired after removing the time-dependent drift of the orbits.

The assumption that the beam shapes are factorisable into independent transverse proton density functions of x and y is tested, and the potential bias is estimated using specific (imaging, offset and diagonal) scans or by studying the parameters of the luminous region in standard vdM scans.

The electromagnetic interaction between the two beams leads to an optical distortion effect on the bunch shapes (dynamic-$\beta^*$) and a deflection from the nominal position. These beam-beam effects are corrected for during the calibration process.

Accurate measurements of bunch intensities are necessary to evaluate the visible cross-section estimates. Each filled bunch is composed of roughly 8$-$9$\times$10$^{10}$ protons, as determined by specialized detectors at the LHC. Corrections are applied to account for any spurious charges, including "ghosts" present in nominally empty bunches and "satellites" in neighbouring buckets of the filled bunch.

\paragraph{Integration uncertainties}

\noindent During nominal physics conditions, measurements are influenced by out-of-time signals. These signals are created by the electronic spillover or the de-excitation of the surrounding material after the collisions. Additionally, the luminometer efficiency may also vary over time due to irradiation, aging or other detector-specific effects. To obtain precise luminosity estimates, corrections must be applied for these effects. After applying all necessary corrections, residual stability- and linearity-related uncertainties are attributed to the measurement, determined by a long-term comparison of the measured luminosities with the various counting methods.

\begin{table}[ht]
    \centering
    {\begin{tabular}{@{}lcccc@{}}
    Source                      &   2015 [\%]   &   2016 [\%]   &   2017 [\%]   &   2018 [\%]   \\
    \hline
    \multicolumn{5}{c}{Calibration uncertainties} \\
    Length scale                &   0.2         &   0.3         &   0.3         &   0.2     \\
    Orbit drift                 &   0.8         &   0.5         &   0.2         &   0.1     \\
    Transverse factorizability  &   0.5         &   0.5         &   0.8         &   2.0     \\
    Beam-beam effects           &   0.5         &   0.5         &   0.6         &   0.2     \\
    Beam current calibration    &   0.2         &   0.2         &   0.3         &   0.2     \\
    Ghosts and satellites       &   0.1         &   0.1         &   0.1         &   0.1     \\
    Cross-detector consistency  &   0.6         &   0.3         &   0.6         &   0.5     \\
    \multicolumn{5}{c}{Integration uncertainties} \\
    Afterglow                   &   0.3$\oplus$0.1     &   0.3$\oplus$0.3     &   0.2$\oplus$0.3     &   0.1$\oplus$0.4 \\
    Cross-detector stability    &   0.6         &   0.5         &   0.5         &   0.6     \\
    Linearity                   &   0.5         &   0.3         &   1.5         &   1.1     \\
    CMS deadtime                &   0.5         &   $<$0.1      &   0.5         &   $<$0.1  \\
    \hline
    Total uncertainty           &   1.6         &   1.2         &   2.3         &   2.5     \\
    \end{tabular}
    \caption{List of the contributions to the systematic uncertainty in each year of Run~2~\cite{CMS:LUM-17-003,CMS:LUM-17-004,CMS:LUM-18-002}. The values for 2017 and 2018 are preliminary.}
    \label{uncertainties}}
\end{table}

Luminosity measurement precision at the CMS experiment reached 1.6\% uncertainty for the combination of the full Run~2 13 TeV pp data taking period. The precisions for the 2017 and 2018 data-taking periods are expected to be reduced after the improved final analysis of the dominant systematic sources.

{\let\thefootnote\relax\footnote{This work was supported by the National Research, Development and Innovation Office of Hungary (K~143460, K~146913, K~146914).}}

\bibliographystyle{JHEP}
\bibliography{lumi_lhcp_bib}

\end{document}